\documentclass{llncs}
\usepackage[utf8]{inputenc}
\usepackage{mathtools}
\usepackage{hyperref}
\usepackage{bbm}
\usepackage{amsmath}
\usepackage{amssymb}
\usepackage{cite}
\usepackage{tikz}
\usetikzlibrary{arrows,automata, positioning}

\usepackage{fancyhdr}
\usepackage{lastpage}
\usepackage{listings}
\usepackage{xcolor}
\lstdefinestyle{padrao}{
basicstyle=\scriptsize,
columns=fullflexible,
mathescape=true,
tabsize=2, linewidth=\textwidth, 
numbers=left,
numberstyle=\tiny,
stepnumber=1,
numbersep=5pt,
backgroundcolor=\color{gray!20},
showspaces=false,
showstringspaces=false,
showtabs=false,
frame=lines,
tabsize=2,
captionpos=b,
floatplacement={tbp},
breaklines=true,
breakatwhitespace=false,
escapeinside={\%*}{*)},
numberbychapter=false
}

\lstdefinelanguage{maude}{
 keywords={pr, inc, fmod, endfm, mod, is, endm, tmod, endtm, sort, sorts, subsort, op, ops, eq, rl, ceq, if, crl, assoc, comm, ctor, id, var, vars, mb, cmb,owise,load, reduce, search, nonexec, protecting, Timed, time, and, with, mode, default, increase, No, solution, in}
 }

\lstnewenvironment{maude}[1][]{\lstset{language=maude,style=padrao,#1}}{}

\hypersetup{colorlinks=true,urlcolor=blue,linkcolor=blue,citecolor=blue}
\title{Towards the Modular Specification and Validation of Cyber-Physical Systems}
\subtitle{A Case-Study on Reservoir Modeling with Hybrid Automata}
\author{Andre Metelo\inst{1} \and Christiano Braga\inst{1} \and Diego Brand\~ao \inst{2}}
\institute{Instituto de Computa\c{c}\~ao \\ Universidade Federal Fluminense \\ 
\email{\{metelo,cbraga\}@ic.uff.br}\and Centro Federal de Educa\c{c}\~ao Tecnol\'ogica Celso Suckow da Fonseca\\ CEFET-RJ \\ \email{diego.brandao@eic.cefet-rj.br}}
\begin{document}

\maketitle

\begin{abstract}

Cyber-Physical Systems (CPS) are systems controlled by one or more computer-based components tightly integrated with a set of physical components, typically described as sensors and actuators, that can either be directly attached to the computer components, or at a remote location, and accessible through a network connection. The modeling and verification of such systems is a hard task and error prone that require rigorous techniques.  Hybrid automata is a formalism that extends finite-state automata with continuous behavior, described by ordinary differential equations. This paper uses a rewriting logic-based technique to model and validate CPS, thus exploring the use of a \emph{formal} technique to develop such systems that combines  expressive specification with efficient state-based analysis. Moreover, we aim at the \emph{modular} specification of such systems such that each CPS component is \emph{independently} specified and the final system emerges as the \emph{synchronous product} of its constituent components.  
We model CPSs using Linear Hybrid Automaton and implement them in Real-Time Maude, a rewriting logic tool for real-time systems. With this method, we develop a specification for the $n$-reservoir problem, a CPS that controls a hose to fill a number of reservoirs according to the physical properties of the hose and the reservoirs. 


\end{abstract}

\section{Introduction}\label{sec:intro}

Cyber-Physical Systems (CPS)~\cite{CPSPrinciples} are ever present in our daily life. They can be intuitively described as systems that are controlled by one or more computer based components tightly integrated with a set physical components, typically described as sensors and actuators that can either be directly attached to the computer components, or at a remote location and accessible through a network connection. 

Most CPS have to cope with design requirements that are imposed 
onto them by their multiple applications in the real world. Typically a CPS has to be specified and tested against environments that require the system to: 
\begin{itemize}
\item operate in real-time,
\item realize reactive computations,
\item leverage concurrent and distributed processing,
\item deal with synchronization issues.
\end{itemize}


In ~\cite{CPSPrinciples}, one of the major books on CPS in a \emph{vast} (e.g. ~\cite{10.1007/3-540-57318-6_30,Henzinger2000, Bu:2011,Zhang:2017,Thomas:1991:AIO:114891.114895,shafi:2012,ALUR1994183,reservoiarprobelm}) literature on the subject, Alur describes how Linear Hybrid Automata (LHA) can be used for modeling CPS. In this context, the  $2$-reservoirs problem~\cite{reservoiarprobelm}, a text-book problem on dynamic systems where a control system needs to decide to which of two tanks a hose needs to be moved given the reservoirs and hose's physical characteristics, is a CPS and therefore can be modeled as a LHA. In this paper we \emph{generalize} this problem to an \emph{arbitrary} number of reservoirs, each with their \emph{individual} physical characteristics, and by adding \emph{latency} to hose dislocation. We model and analyze both the standard problem description and the generalized version using Rewriting Logic~\cite{meseguer:1992}, an expressive formalism for the specification and verification of concurrent and distributed systems~\cite{olveczky2018designing}. Moreover, we specify the $n$-reservoir system \emph{modularly} as the \emph{synchronous product}~\cite{Arnold:1994:FTS:176787} of its constituent components.  

This paper contribution is manifold 
: (i) a precise definition of the synchronous product of real-time rewrite systems, extending~\cite{narciso:2016}, (ii) a model of the $n$-reservoir problem as an LHA, (iii) how to describe a CPS as a LHA in Rewriting Logic by representing its components, sensors, actuators and controllers, as mathematical tuples denoting objects that communicate asynchronously, (iv) a modular specification of the $n$-reservoir system, based on (i), and (v) its implementation and model checking in Real-Time Maude (RTM)~\cite{RTMaude}, a Rewriting Logic tool designed for the formal specification and analysis of real-time and hybrid systems. This is a first-step in the development of a formal method and its tooling to \emph{modularly} specify and verify Cyber-Physical Systems based on its Rewriting Logic semantics of the associated Linear Hybrid Automata.


The remainder of this paper is organized as follows.
Section~\ref{sec:prem} describes foundational requirements related to LHA and RTM. 
In Section~\ref{sec:reservoir-RTM}, we discuss our prototype implementation of the reservoir problem in RTM. Section~\ref{sec:rt-syncprod} presents the synchronous product of rewrite systems and its extension to real-time, together with the modular specification of the $2$-reservoir problem.
Section~\ref{sec:relatedwork} describes related work. Section~\ref{sec:conclusion} concludes this paper describing the insights and future research based on our findings.  

\section{Preliminaries}\label{sec:prem}
In oder to develop the implementation of the $n$-reservoirs problem and its model check through the use of Linear Hybrid Automaton and Real-Time Maude an understanding of what they are is required. This section provides a basic introduction of these topics.

\subsection{Linear Hybrid Automaton (LHA)}
A LHA is a Finite State Machine that is associated with a finite set of variables that are described by ordinary differential equations (ODE). To guarantee that the solutions of the differential equation are well defined, we assume that the ODE are Lipschitz continuous \cite{reservoiarprobelm}. Moreover, these differential equations are such that any test and attribution within the model of the LHA are affine, that is, a linear equation of the form \(a_1x_1+a_2x_2+...+a_nx_n \sim 0\) where \(\sim\) is a comparison operation that can be one of \(<, \le, =, \ge or >\) and an attribution is in the form \(x_i = a_o + a_1x_1+a_2x_2+...+a_nx_n\) and \(a_0, a_1, ..., a_n\) are integer or real constants. 

A LHA \textbf{HP} consists of:
\begin{enumerate}
\item An asynchronous process \textbf{P}, where some of its state variables can be of type cont, and appear only in affine tests and affine assignments in the guards and updates of the tasks of \textbf{P};
\item A continuous-time invariant \textbf{CI}, which is a Boolean expression over the state variables \textbf{S}, where the variables are continuous (that is, of type cont) and appear only in affine tests;
\item A rate constraint \textbf{RC}, which is a Boolean expression over the discrete state variables and the derivatives of the continuously updated state variables that appear only in affine tests.
\end{enumerate}

Inputs, outputs, states, initial states, internal actions, input actions, and output actions of the LHA \textbf{HP} are the same as that of the asynchronous process \textbf{P}. Given a state \textbf{s} and a real-valued time \(\delta > 0\), \(s \xrightarrow{\delta}\ s + \delta r\) is a timed action of \textbf{HP}, for a rate vector \textbf{r} consisting of a constant \(r_x\) for every continuously updated variable $x$ if:
\begin{enumerate}
\item The expression \textbf{RC} is satisfied for every continuously updated variable $x$, the derivative \(\dot{x}\) is assigned the value \(r_x\) and every discrete variable $x$ is assigned the value \(s(x)\);
\item The state \(s + tr\) satisfies the expression \textbf{CI} for all values \(0 \leq t \leq \delta\).
\end{enumerate}

\begin{figure}
\centering
\scalebox{0.7}{
\begin{tikzpicture}[->,>=stealth',shorten >=1pt,auto, node distance=6.5cm, semithick]   
\node[state, inner sep=-0.1cm] (q1) 
	{$\begin{array}{c}
		\mathit{State}~1\\ \\
		\mathit{Variables~constraints} \\
		\mathit{Assignments}
	    \end{array}$};    
\node [left = 2cm of q1] (init) {} ;
\node[state, inner sep=-0.1cm] (q2) [right of=q1] 	
	{$\begin{array}{c}
		\mathit{State}~2\\ \\
		\mathit{Variables~constraints} \\
		\mathit{Assignments}
	    \end{array}$};
\path (init) edge [above] node {$\mathit{Initialization}$} (q1) ; 
\path (q1) edge node {$\mathit{Test} \quad \mathit{Assignment}$} (q2) ; 
\end{tikzpicture}
}
\caption{Simple extended state machine diagram of a LHA}
\label{pic:statemachine}
\end{figure}
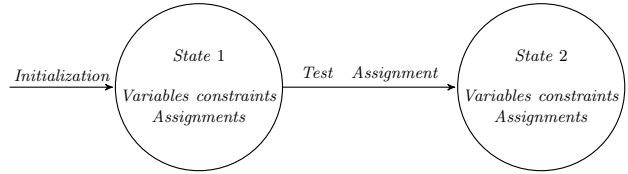

As such, a LHA can be represented as an extended state machine as shown in Figure~\ref{pic:statemachine}, where the Initial Variables represent the starting values of all constants and discrete variables in the LHA; State($i$), $i \in \{1,2\}$, represents one of the many states in LHA together with tests and assignments that occur while the system is evolving in time while not changing state; the arrow with a test and attribution represents the boolean test that needs to be satisfied for the transition between one state and another to happen, alongside any changes that must be assigned to LHA variables.

\subsection{Real-time rewrite systems and Real-Time Maude}\label{sec:rtm}

A Rewriting Logic theory is essentially a triple $(\Sigma, E, R)$ where $\Sigma$ is a typed (or sorted) signature (many sorted, order-sorted or membership equational, that is, Rewriting Logic is \emph{parameterized} by a choice of equational logic), $E$ is a set of $\Sigma$-equations and $R$ is a set of $\Sigma/E$-rules where the terms being rewritten are those in the initial $\Sigma$-algebra $\mathcal{T}_\Sigma$ identified by $\Sigma$-equations $E$.

A real-time rewrite theory is a tuple $R = ( \Sigma, E \cup A, R)$~\cite{olveckzy:2002}, where: (i) $(\Sigma, E \cup A)$ contains an equational subtheory $(\Sigma_{\mathit{TIME}}, E_\mathit{TIME}) \subseteq (\Sigma , E \cup A)$, satisfying the \textit{TIME} axioms that specifies sort \texttt{Time} as the time domain (which can be discrete or dense) and $\{\_\}$ is a built-in constructor of sort \texttt{GlobalSystem}.  
The rules in $R$ are decomposed into: (i) instantaneous rewrite rules, that do not act on the system as a whole, but only on some system components, and (ii) tick rules that model the elapse of time in a system, having the form $l : \{t\} \stackrel{u}{\longrightarrow} \{t'\} ~\mathit{if}~\mathit{condition}$, where $t$ and $t'$ are terms of sort \texttt{System}, $u$ is a term of sort \texttt{Time} denoting the duration of the rewrite. 
Given a real-time rewrite theory $R$, a \emph{computation} is a non-extensible sequence $t_0 \longrightarrow t_1 \longrightarrow \ldots \longrightarrow t_n$ (that is, one for which $t_n$ cannot be further rewritten) or an infinite sequence $t_0 \longrightarrow t_1 \longrightarrow \ldots$ of one-step $R$-rewrites $t_i \longrightarrow t_i +1$, with $t_i$ and $t_i+1$ ground terms, starting with a given initial term $t_0$ of sort \texttt{System}. 

Maude~\cite{Maude} is a system/language that implements 
concurrent systems through the use of equations and rewrite rules specified in one or more modules. 
Maude itself has been extended through Full Maude, which is fully written in Maude itself, to add several features to the system/language. These features include, but are not limited to, object-oriented modules, module parameterization, and $n$-tuple declaration. 

Real-Time Maude (RTM) is an extension of Full Maude that includes the requirements and tooling to model and check real-time systems. The time evolution is achieved through tick rules that determine the effects of time in the system. As such, RTM allows for a very granular control over how a system can evolve both in time and instantaneously by means of two classes of rewrite rules that specify either timed or discrete transitions. 
Although a timed module is parametric on the time domain, Real-Time Maude provides some predefined modules specifying useful time domains. For example, the modules \texttt{NAT-TIME-DOMAIN-WITH-INF} and \texttt{POSRAT-TIME-DOMAIN-WITH-INF} define the time domain to be, respectively, the natural numbers and the nonnegative rational numbers, and contain the subsort declarations \texttt{Nat < Time} and \texttt{Pos-Rat < Time}.
In Real-Time Maude, tick rules, together with their durations, are specified using the syntax
\texttt{crl [l] : \{t\} => \{t'\} in time u if condition}.

Essentially, an RTM specification representing a hybrid automaton $\mathcal{H} = (S, \to_t, \to_d)$, with $S$ the set of states of $\mathcal{H}$, $\to_t$ the set of timed transitions, and $\to_d$ the set of discrete transitions, is given by a structure $(\Sigma, E, R_t, R_d)$ where the equational specification $(\Sigma, E)$ specifies the set $S$ of states of the hybrid automaton $\mathcal{H}$, $R_t$ is the set of $\Sigma/E$-timed rewrite rules representing $\to_t$ transitions in $\mathcal{H}$, and $R_d$ the set of  $\Sigma/E$-discrete rewrite rules representing $\to_d$ transitions in $\mathcal{H}$.

We describe RTM specification language in Section~\ref{sec:reservoir-RTM}, by example, while describing the implementation of the $2$-reservoir problem. 

\section{The reservoir problem in RTM}\label{sec:reservoir-RTM}

The $2$-reservoir problem was presented by John Lygeros \emph{et al.} in \cite{reservoiarprobelm}. The problem has been fully defined in that work, alongside its differential equations and its LHA. 

\begin{figure}
\centering
\includegraphics[scale=0.4]{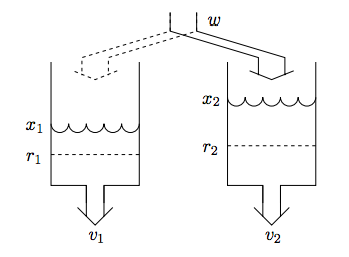}
\caption{$2$-reservoirs Diagram \cite{reservoiarprobelm}}
\label{pic:diagram}
\end{figure}

Succinctly, the problem has as a pair of reservoirs that are flowing water out of the system at a constant rate. Water is added to the system through a hose that has a constant intake rate. The hose can be moved from one reservoir to the other instantaneously. 
There is a control system that is designed to make sure that the water level in each reservoir does not fall below a predefined level. This can be seen in Figure~\ref{pic:diagram}.

\begin{figure}
\centering
\scalebox{0.75}{
\begin{tikzpicture}[->,>=stealth',shorten >=1pt,auto,node distance=5cm, semithick]
\node (start1) {} ;
\node[state] (q1) [below right of=start1] 
	{$\begin{array}{c}
		q_1\\ 
		\stackrel{\cdot}{x_1} = w - v_1 \\
		\stackrel{\cdot}{x_2} = - v_2 \\ \\
		x_2 \ge r_2		
	    \end{array}$};
\node[state] (q2) [right of=q1] 	
	{$\begin{array}{c}
		q_2\\ 
		\stackrel{\cdot}{x_1} = - v_1 \\
		\stackrel{\cdot}{x_2} = w - v_2 \\ \\
		x_1 \ge r_1		
	    \end{array}$};
\node (start2) [above right of=q2] {} ;
\path [shorten <= 2cm] (start1) edge [above] node {$x_1 \ge r_1 \land x_2 \ge r_2$} (q1) ; 
\path [shorten <= 2cm] (start2) edge [above] node  {$x_1 \ge r_1 \land x_2 \ge r_2$} (q2) ; 
\path (q1) edge [bend left] node {$x_2 \ge r_2 \quad x' := x$} (q2) ; 
\path (q2) edge [bend left] node {$x_1 \ge r_1 \quad x' := x$} (q1) ; 
\end{tikzpicture}
}

\caption{$2$-reservoir LHA Diagram}
\label{pic:reservoirLHA}
\end{figure}
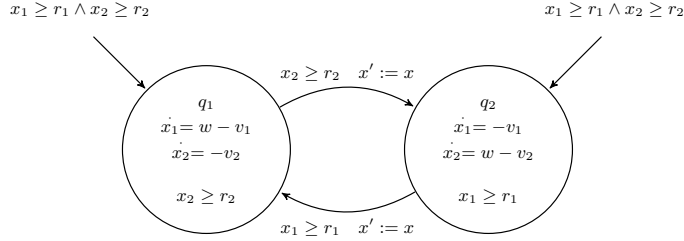

The LHA of the $2$-reservoirs problem is described in Figure~\ref{pic:reservoirLHA}, where: 
\(x_i\) is the water level at the reservoir $i$ (that is time dependent), 
\(r_i\) is the flow of water out of the reservoir $i$,
\(q_i\) is the state of the system with the hose filling reservoir $i$,
and \(w\) is the hose's water flow rate.

\paragraph{The generalized $n$-reservoir problem} is the extension of the $2$-reservoirs problem where the system has a finite non-predetermined number of reservoirs, each with its own individual physical characteristics. In this new structure, the control system has to make the decision to which reservoir it will move the hose to when multiple reservoirs are potentially below the minimum water level threshold.

Although the differential equations of each individual reservoir stay the same, the total number of continuous variables that are tracked by the system scales with the size of the instance (\(n\) continuous variables). Additionally, the number of states in the LHA grows with the number of reservoirs as well (\(2n\) states). Similarly, the problem will have a considerably larger number of state transitions. These extra states and transitions force the control system to cope with a more complex decision process to choose where to move the hose at any specific point in time (\(n(n-1)\) transitions).

We are now ready to inspect how the reservoirs problems can be implemented in RTM with LHA in place and a clear understanding of the differences between the simple (with $2$ reservoirs) and the generalized versions of the reservoir problem.

\subsection{The $2$-reservoir problem in RTM}

The states of the hybrid automaton modeling the $2$-reservoir problem is defined in RTM through a triple $\langle \{\mathit{right}, \mathit{left}\}, \mathbbm{Q}^+, \mathbbm{Q}^+ \rangle$ where $\{\mathit{right}, \mathit{left}\}$ denotes the hose position (either on the right or left reservoir) and $\mathbbm{Q}^+$ is the set of positive rational numbers denoting the height of the water column for each reservoir. 

The triple and constants can be defined through the code below 
where keywords \emph{op} and \emph{ops} declare operations for the left and right position of the hose, the system configuration (that is, the triple $\langle \{\mathit{right}, \mathit{left}\}, \mathbbm{Q}^+, \mathbbm{Q}^+ \rangle$) and constants for the hose flow rate $w$, each reservoir flow rate $v_i$, and the water level on each reservoir $r_i$, with $i \in \{1,2\}$, and \texttt{NNegRat} is the sort for $\mathbbm{Q}^+$, respectively.   
\begin{maude}[caption=Signature for the states of the $2$-reservoir hybrid automaton, label=lst:2-res-sig]
ops left right : -> Hose [ctor] .
op _`,_`,_ : Hose NNegRat NNegRat -> System [ctor] .
ops w v1 v2 r1 r2 : -> NNegRat .
\end{maude} 

The movement of the hose can be represented by rules that change the system configuration. Rule \emph{moveright} moves the hose from left to right, while \emph{moveleft} goes the other way around. The rules are triggered when the water level of the proper reservoir reaches the minimum acceptable level.
\begin{maude}
vars x1 x2 : NNegRat .
crl [moveright] : left,x1,x2 => right,x1,x2 if x2 <= r2 . 
crl [moveleft] : right,x1,x2 => left,x1,x2 if x1 <= r1 .
\end{maude} 

The last step is to create the rules specifying how water levels change in time, the so called \emph{tick} rules. The water levels increase in linear time by $w - v_i$ and decreases, also in linear time, by $v_i$.

\begin{maude}
var R : Time .
crl [tick-right] :
	{right, x1, x2} => {right, x1 - (v1 * R), x2 + ((w - v2) * R)} in time R 
if x1 > r1 [nonexec] .
			
crl [tick-left] :
	{left, x1, x2} => {left, x1 + ((w - v1) * R), x2 - (v2 * R)} in time R
if x2 > r2 [nonexec] .
\end{maude} 

The evolution in time does not happen when the water level of the reservoir with the hose is above the water threshold and the second reservoir is below or at its water threshold. Without this restriction, the system can evolve to undesired configurations where it keeps filing a reservoir that is above minimum threshold while letting the other reservoir dry out.

\subsection{The $n$-reservoir problem in RTM}

The implementation of the $n$-reservoir problem requires a dynamic structure to accommodate the run-time defined reservoir count. For that we leverage Maude's sets to create a \emph{System} configuration made of a hose and each individual reservoir as shown below.
\begin{maude}
{ hose(10, 0) < 0 | thr: (15, 50), hth: 30, rte: 5 > < 1 | thr: (15, 50), hth: 30, rte: 5 > }
\end{maude}

In this structure we have a hose element that has an in-take rate of $10$ units per time unit and it is positioned on top of reservoir $0$. The hose is followed by multiple structures in the form \(<\) 0 \(|\) thr: (15, 50), hth: 30, rte: 5 \(>\) that represent a single reservoir with its unique identifier, the upper and lower water thresholds, the current water level and the constant flow of water out of the reservoir.

The hose is defined by:
\begin{maude}
sort Hose .
op hose : NNegRat Nat -> Hose [format(m! o)] .
\end{maude}

A single reservoir, logically defined as \(<\) N \(|\) List of attributes \(>\), can be coded as:
\begin{maude}
subsort ReservoirAttribute < ReservoirAttributes .
op <_|_> : Nat ReservoirAttributes -> Reservoir [ctor format(b! o b! o b! o)] .
op _`,_ : ReservoirAttributes ReservoirAttributes -> ReservoirAttributes [ctor assoc comm] . 
\end{maude} 

Then the reservoir attributes are defined.
\begin{maude}
*** Upper and Lower Water thresholds
op thr: : NNegRat NNegRat -> ReservoirAttribute [ctor format(b! o)] . 
*** Water Level
op hth:_ : NNegRat -> ReservoirAttribute [ctor format(b! o b!)] . 
*** Leak Rate
op rte:_ : NNegRat -> ReservoirAttribute [ctor format(b! o b!)] . 
\end{maude}

The whole system is put together by concatenating a Hose and as many reservoirs as desired. 
\begin{maude}
subsort Hose Reservoir < System .
op __ : System System -> System [ctor assoc comm] .
\end{maude} 

Keep in mind that in Maude, computations are identified with rewritings. One of the distinguished features of Maude is to implement rewriting modulo axioms, such as associativity and commutativity. 
Thus the use of definition of many \emph{assoc} and \emph{comm} in the operators definition.

The next step is to define functions - done through operators in Maude - to fill and drain the reservoirs alongside tests to validate that the evolution in time can occur. First, a function to add water to the reservoir that the hose is currently pointing to:
\begin{maude}
*** Vars used across the functions and rules
vars N M : Nat .
vars L U Ln Un Lm Um Xn Xm Dn Dm W : NNegRat . 
var T : Time . var S : System . 
vars RA RAn RAm : ReservoirAttributes .

op fill : Reservoir NNegRat NNegRat -> Reservoir .
eq fill(< N | thr: (L, U) , hth: Xn , rte: Dn >, W, T) = 
	< N | thr: (L, U), hth: (Xn + ((W - Dn) * T)), rte: Dn > .
\end{maude} 

Next, a function that drains water from all reservoirs:
\begin{maude}
op drain : Reservoir NNegRat -> System .
eq drain(< N | thr: (L, U), hth: Xn, rte: Dn >, T) = 
	< N | thr: (L, U), hth: sd(Xn,(Dn * T)), rte: Dn > .
eq drain(< N | thr: (L, U), hth: Xn, rte: Dn > S, T) = 
	< N | thr: (L, U), hth: sd(Xn,(Dn * T)), rte: Dn > drain(S, T) .
\end{maude} 
and a test to identify if there is a reservoir that needs refill:
\begin{maude}
op refill? : System -> Bool .
ceq refill?(< N | thr: (L, U), hth: Xn, RAn >) = true if Xn <= L .
ceq refill?(< N | thr: (L, U), hth: Xn, RAn > S) = true if Xn <= L .
eq refill?(S) = false [owise] . 
\end{maude} 

Next, the rules to control the movement of the hose and the evolution in time must be defined. A single rule is capable of moving the hose from a reservoir above the lower threshold to one that is below the lower threshold. If there is no reservoir below the lower threshold, or the current reservoir is below the threshold, the hose stays in the same place. 

\begin{maude}
crl [move-hose] : 
	hose(W, N) < N | thr: (Ln, Un), hth: Xn, RAn > 
	< M | thr: (Lm, Um), hth: Xm, RAm > S =>
	hose(W, M) < N | thr: (Ln, Un), hth: Xn, RAn >
	< M | thr: (Lm, Um), hth: Xm, RAm > S
if (Xm <= Lm) and (N =/= M) and (Xn >= Ln) . 
\end{maude} 

In rule \emph{move-hose}, the configuration \texttt{ hose(10, 0) \(<\) 0 \(|\) thr: (15, 50), hth: 40, rte: 5 \(>\) \(<\) 1 \(|\) thr: (15, 50), hth: 15, rte: 5 \(>\) }  changes to configuration \texttt{ hose(10, 1) \(<\) 0 \(|\) thr: (15, 50), hth: 40, rte: 5 \(>\) \(<\) 1 \(|\) thr: (15, 50), hth: 15, rte: 5 \(>\) } as the reservoir $1$ \emph{hth} attribute value is less or equal to the lower threshold defined in the \emph{thr} attribute. 

Once the system is in a configuration with no reservoir below the lower threshold or if there is at least one reservoir that needs water and the hose is placed on one of them, the system can evolve in time. 

\begin{maude}
crl [tick] : {hose(W, N) < N | RAn > S} => 
	{hose(W, N) fill(< N | RAn >, W, T) drain(S, T)} in time T 
if not refill?(S) [nonexec] .
\end{maude} 

In this rule, the system \texttt{ hose(10, 1) \(<\) 0 \(|\) thr: (15, 50), hth: 40, rte: 5 \(>\) \(<\) 1 \(|\) thr: (15, 50), hth: 15, rte: 5 \(>\) } can evolve to \texttt{ hose(10, 1) \(<\) 0 \(|\) thr: (15, 50), hth: 30, rte: 5 \(>\) \(<\) 1 \(|\) thr: (15, 50), hth: 20, rte: 5 \(>\) } while the system \texttt{ hose(10, 0) \(<\) 0 \(|\) thr: (15, 50), hth: 40, rte: 5 \(>\) \(<\) 1 \(|\) thr: (15, 50), hth: 15, rte: 5 \(>\) } is not eligible to evolve in time because there is a reservoir below the lower threshold, and the hose is on a reservoir above the threshold.

Maude offers techniques to search for states that are reachable from the initial states and match a given search pattern. 
To demonstrate this property, we search the first six steps of the system evolution starting with initial state given by \texttt{\(<\) 0 \(|\) hth: 30, rte: 5, thr:(15,50)\(>\) \(<\) 1 \(|\) hth: 30, rte: 5, thr:(15, 50)\(>\) \(<\) 2 \(|\) hth: 30, rte: 5, thr:(15,50)\(>\)}.

\begin{maude}
Timed search in TEST 
	{init2} =>* {S:System} 
in time < 5 and with mode default time increase 1 :

Solution 1
S:System --> < 0 | hth: 30,rte: 5,thr:(15,50)> < 1 | hth: 30,rte: 5,thr:(15,
    50)> < 2 | hth: 30,rte: 5,thr:(15,50)> hose(10,0); TIME_ELAPSED:Time --> 0

Solution 2
S:System --> < 0 | hth: 35,rte: 5,thr:(15,50)> < 1 | hth: 25,rte: 5,thr:(15,
    50)> < 2 | hth: 25,rte: 5,thr:(15,50)> hose(10,0); TIME_ELAPSED:Time --> 1

Solution 3
S:System --> < 0 | hth: 40,rte: 5,thr:(15,50)> < 1 | hth: 20,rte: 5,thr:(15,
    50)> < 2 | hth: 20,rte: 5,thr:(15,50)> hose(10,0); TIME_ELAPSED:Time --> 2

Solution 4
S:System --> < 0 | hth: 45,rte: 5,thr:(15,50)> < 1 | hth: 15,rte: 5,thr:(15,
    50)> < 2 | hth: 15,rte: 5,thr:(15,50)> hose(10,0); TIME_ELAPSED:Time --> 3

Solution 5
S:System --> < 0 | hth: 45,rte: 5,thr:(15,50)> < 1 | hth: 15,rte: 5,thr:(15,
    50)> < 2 | hth: 15,rte: 5,thr:(15,50)> hose(10,1); TIME_ELAPSED:Time --> 3

Solution 6
S:System --> < 0 | hth: 45,rte: 5,thr:(15,50)> < 1 | hth: 15,rte: 5,thr:(15,
    50)> < 2 | hth: 15,rte: 5,thr:(15,50)> hose(10,2); TIME_ELAPSED:Time --> 3

No more solutions
\end{maude}

Next, in order to estimate RTM's efficiency when executing in entry-level computer devices, we executed a search looking for a specific system configuration from the \emph{init2}  configuration.  

    
\begin{maude}
Timed search in TEST 
	{init2} =>* {< 0 | hth: 45,RA0:ReservoirAttributes > < 1 | hth: 10,
    RA1:ReservoirAttributes > < 2 | hth: 10,RA2:ReservoirAttributes >} 
in time < 100 and with mode default time increase 1 :

No solution

\end{maude}

In order to make sure we looked through the full state space, a state that the system will not reach was selected, and as
Table \ref{tab:search} demonstrates the whole process took $10$ms of processing and executed over $7,000$ rewritings to accomplish the task.

\begin{table}[htb]
\caption{Evaluation of Search Technique in Maude}
\label{tab:search}
\centering
\begin{tabular}{|c|c|c|c|}
\hline
Test & Time (in ms) & Rewrites &  Rewrites/second\\
\hline
$1$ & $10$ & $7978$ & $730786$\\
$2$ & $10$ & $7029$ & $692580$\\
\hline
\end{tabular}
\end{table}

\subsection{Model Checking}
An RTM specification induces a timed automaton. Standard Maude model checking techniques may be applied to validate timed Linear Temporal Logic formula over a given timed automaton. This is achieved in RTM through a module extension that includes the \emph{TIMED-MODEL-CHECK} module into the reservoir problems implemented. 

With the \emph{TIMED-MODEL-CHECK} in a new RTM module, it is possible to verify if the system evolves to a point where it has reservoirs that are below the lower water threshold, and need the hose to be moved to - as a temporal logic: \(\models t \: \square \: \lnot \lozenge \: one\mbox{-}down\), where \emph{one-down} represents the event of at least one reservoir below the lower water threshold. The second test executed verified if, at any point in time, all reservoirs need are below the lower treshold - as a temporal logic proposition: \(\models t \lnot \square \: \lozenge \: macondo\), where \emph{macondo} represents the event of all reservoirs reaching the lower threshold at any point in time.

If the system is well-formed - where \(\sum_{i=1}^{n} rte_{i} = w\) - the checker should not find a state where all reservoirs reach their low threshold or their upper threshold. During the execution of such system up to $n-1$ reservoirs could cross below the low threshold. This behavior is achieved after the model check execution. It can not find a solution where all reservoirs fall below the lower threshold; and it produces a counter example when asked if we can identify a state where the the water of a least one reservoir drops to or below the lower threshold.
\begin{maude}
Model check{init2}  |=t ~[]<> macondo in TEST in time < 
	5 with mode default time increase 1
Result Bool :
  true

Model check{init2}  |=t[]~ <> one-down in TEST in time < 5
	with mode default time increase 1
Result ModelCheckResult :
  counterexample(
	{{	< 0 | hth: 30,rte: 5,thr:(15,50)> < 1 | hth: 30,rte: 5,thr:(15,50)> 
        < 2 | hth: 30,rte: 5,thr:(15,50)> hose(10,0)} in time 0,'tick}
  	{{	< 0 | hth: 35,rte: 5,thr:(15,50)> < 1 | hth: 25,rte: 5,thr:(15,50)>
        < 2 | hth: 25,rte: 5,thr:(15,50)> hose(10,0)} intime 1,'tick}
    {{	< 0 | hth: 40,rte: 5,thr:(15,50)> < 1 | hth: 20,rte: 5,thr:(15,50)>
        < 2 | hth: 20,rte: 5,thr:(15,50)> hose(10,0)} in time 2,'tick}
    {{	< 0 | hth: 45,rte: 5,thr:(15,50)> < 1 | hth: 15,rte: 5,thr:(15,50)>
        < 2 | hth: 15,rte: 5,thr:(15,50)> hose(10,0)} in time 3,'move-hose}
    ,{{	< 0 | hth: 45,rte: 5,thr:(15,50)> < 1 | hth: 15,rte: 5,thr:(15,50)> 
        < 2 | hth: 15,rte: 5,thr:(15,50)> hose(10,1)} in time 3,'move-hose}
    {{	< 0 | hth: 45,rte: 5,thr:(15,50)> < 1 | hth: 15,rte: 5,thr:(15,50)> 
        < 2 | hth: 15,rte: 5,thr:(15,50)> hose(10,2)} in time 3,'move-hose})
\end{maude}

Table \ref{tab:modelchecking} demonstrates the amount of time required to complete the model checking of the safety and liveness requirements for the system using the \emph{init2} configuration, with both proofs computed under 100 ms in an entry level x86 system, this indicates how efficient the rewriting system in RTM is at running these LTL based model-checks.
\begin{table}[htb]
\caption{Model checking Evaluation}
\label{tab:modelchecking}
\centering
\begin{tabular}{|c|c|c|c|}
\hline
Test & Time (in ms) & Rewrites & Rewrites/second\\
\hline
$1$ & $12$ & $7074$ &$570529$\\
$2$ & $21$ & $9580$ &$443046$\\
\hline
\end{tabular}
\end{table}

The RTM code for both the $2$-reservoir and the $n$-reservoir modules and the model checking can be retrieved from~
\url{https://github.com/andremetelo/CPSSources/tree/master/Reservoir}.

\subsection{Preliminary Analysis of LHA and RTM for CPS Design}
A criteria to review a CPS design process has to turn the CPS characteristics into key metrics in its evaluation. The control logic should be embedded in the specification itself. The same should apply for sensors and actuators. Considering that reactive computing and synchronization are intrinsic characteristics of a CPS, the computing model process must naturally support and handle them. Moreover, the time dependency of the differential equations are a constant that has to be managed by the formalization process in order to generate a precise model that represents the physical processes managed by the CPS.

The RTM code used in both variations of the reservoir problem is elegant. It follows a very logical and straight forward principle that satisfies the statements above. Implementing the LHA from figure \ref{pic:reservoirLHA} does not pose a challenge in RTM. The model check tools provided by RTM proved to be straight forward once the tests have been formalized in temporal logic. 

However, the choice of using LHA as the formalization tool imposes a limitation.  A LHA is limited to problems that falls into ordinary differential equations in respect to time. Additionally, it must only use affine tests and attributions. Although many problems that translate into a CPS can fit this model, there are problems that do not fit these requirements.  

\section{The synchronous product of rewrite systems and real-time rewrite systems}\label{sec:rt-syncprod}

The synchronous product of two systems is a procedure to compose such systems such that they evolve simultaneously if they \emph{synchronize} in a given step. A key concept in the synchronous product is the compatibility relation, denoted by $\approx$. Two states, $s_1$ and $s_2$, synchronize on a given \emph{action} if they are \emph{compatible}, that is, iff for each property (or atomic proposition) $p$ shared by both states $p$ holds in $s_1$ iff $p$ holds in $s_2$. 

In~\cite{narciso:2016}, the authors specify that given two rewrite systems $\mathcal{R}_i = (\Sigma_i, E_i \cup A_i, R_i)$, for $i = 1, 2$, their synchronous product, denoted $\mathcal{R}_1 \parallel \mathcal{R}_2$, is a new rewrite system $\mathcal{R} = (\Sigma, E \cup A,R)$ as follows, 
(i)  $\Sigma = \Sigma_1\uplus \Sigma_2 \uplus \Sigma'$, where $\uplus$ denotes the disjoint union of two ets, $\Sigma'$ contains, among other declarations,
 (a) a declaration for the operator \texttt{R.|= : R.State $\times$ R.Prop $\to$ R.Bool}, where notation \texttt{R.\emph{Sort}}, or  \texttt{R.\emph{op}}, denotes sort \texttt{\emph{Sort}}, or operation \texttt{\emph{op}}, from rewrite system \texttt{R};
(b) a declaration for the predicate \texttt{R.$\approx$ : R$_1$.State $\times$ R$_2$.State $\to$ R.Bool}. 
(ii) $E = E_1 \uplus E_2 \uplus E'$, where $E'$ contains, among other declarations:
(a) equations to reduce $s_1 \approx s_2$ to \texttt{true}; 
(iii) $A = A_1 \uplus A_2$;
(iv) $R$ is composed of the following set of rules:
(a) for each rule label $l$ that exists in both systems, say $[l] s_i \to s'_i \in R_i$, we have in $R$ the conditional rule $[l] \langle s1, s2 \rangle \to \langle s'_1, s'_2\rangle \mathit{if} s'_1 \approx s'_2$, with constructor $\langle\_,\_\rangle$ denoting the product state;
(b) for each rule label $l$ that exists in $R_1$ but not in $R_2$, say $[l] s_1 \to s'_1 \in R_1$, we have in $R$ the conditional rule $[l] \langle s_1, x_2\rangle \to \langle s'_1, x_2\rangle ~\mathit{if}~ s'_1 \approx x_2$ (with $x_2$ a variable of sort \texttt{R$_2$.State});
(c) correspondingly for rule labels in $R_2$ but not in $R_1$.

Now in this paper we propose that given two real-time rewrite systems $\mathcal{R}_i = (\Sigma_i, E_i \cup A_i, R_i \cup R_{t_i})$ with $R_i$ instantaneous rewrite rules and tick rules $R_{t_i}$, with the same subequational theory for \textit{TIME},
the synchronized real-time system $\mathcal{R} = (\Sigma, E \cup A, R \cup R_t)$ 
has each individual component composed as above but each tick rule in $R_t$ requires its left-hand side and right-hand side to be $R$-\textit{compatible}, that is, each $r$ in $R_t$ relate product states that are compatible using the rules in $R$. Therefore, a (possibly infinite, Zeno) computation $t_0 \longrightarrow t_1 \longrightarrow \ldots$ in the synchronized real-time system $\mathcal{R}$ has a step $\langle s_{1_i}, s_{2_l}\rangle \stackrel{u}{\longrightarrow} \langle s_{1_j}, s_{2_k}\rangle$ such that the product state $\langle s_{1_j}, s_{2_k}\rangle$ is reached from $\langle s_{1_i}, s_{2_l}\rangle$ after a computation composed by $R$-compatible untimed transitions, that is, transitions resulting from the application of rules in $R$ where each state in the computation is compatible with its predecessor.   

In what follows, we specify (a simplified version) of a \emph{modular} $2$-reservoir problem using the synchronous product extension of Full-Maude available in~\url{http://maude.sip.uc.es/syncprod}. (An extension of Real-Time Maude with real-time synchronous product as described above is under development.) In Listing~\ref{lst:modular-res} we specify the behavior of a reservoir as a component whose state is either \texttt{below} (its fluid threshold) or \texttt{ok}. Whenever the clock ticks, modeled by rule labeled \texttt{tick}, the reservoir (actually, \emph{all} of them as they will synchronize with this action) may change its state from \texttt{ok} to \texttt{below}. And by (re)filling it, it may change from \texttt{below} to \texttt{ok}. State predicate \texttt{refill1?} is true when the state of the reservoir is \texttt{below}.
\begin{maude}[caption=Modular specification of a reservoir, label=lst:modular-res]
(mod RESERVOIR1 is
   including SATISFACTION .  --- declares State, Prop, and |=.
   ops below ok : -> State [ctor] .
   rl [tick] : ok => below .
   rl [fill1] : below => ok .
   op refill1? : -> Prop .
   eq below |= refill1? = true .
   eq ok |= refill1? = false .
endm)
\end{maude}

Now, the $2$-reservoir system is given by the synchronous product of reservoirs $1$ and $2$, as declared in Listing~\ref{lst:modular-2-res} by the statement \texttt{pr RESERVOIR1 || RESERVOIR2 .}. Module \texttt{2-RESERVOIR-SYSTEM} also declares a state proposition specifying that the $2$-reservoir system is safe if reservoirs $1$ and $2$ are not in state \texttt{below} at the same time. 
\begin{maude}[caption=The $2$-reservoir system as the product of reservoirs, label=lst:modular-2-res]
(mod 2-RESERVOIR-SYSTEM is
   pr RESERVOIR1 || RESERVOIR2 .
   op safe : -> Prop [ctor] .
   eq S:State |= safe = not (S:State |= refill1? and S:State |= refill2?) .
endm)
\end{maude}

We may now model check this specification and prove that it is not the case that the system is always safe starting from state \texttt{< ok, ok >} where both reservoirs are above their thresholds. A counter-example is produced showing the  infinite loop \texttt{< ok,ok > $\stackrel{\mathtt{tick}}{\longrightarrow}$ < below,below > $\stackrel{\mathtt{fill2}}{\longrightarrow}$ < below,ok > $\stackrel{\mathtt{fill1}}{\longrightarrow}$ < ok,ok > $\ldots$}.
\begin{maude}[caption=Model checking the $2$-reservoir as a synchronous product, label=lst:mc-2-res-prod]
(mod MODEL-CHECK-2-RESERVOIR-SYSTEM is
   pr 2-RESERVOIR-SYSTEM .
   inc MODEL-CHECKER * (sort State to Conf) .
   inc LTL-SIMPLIFIER .
   subsort Conf < State .
   op init : -> State .
   eq init = < ok, ok > .
endm)

(red modelCheck(init, [] safe) .)
reduce in MODEL-CHECK-2-RESERVOIR-SYSTEM :
  modelCheck(init,[]safe)
result [ModelCheckResult] :
  counterexample({< ok,ok >,'tick},{< below,below >,'fill2}{< below,ok >,
    'fill1}{< ok,ok >,'tick})
\end{maude}

\section{Related work}\label{sec:relatedwork}

Even though CPS are not a new concept, to this day an optimal set of formalisms, languages and tools are not properly defined. In David Broman \emph{et al.} \cite{VPFormLanguages} describe an extensive list of formal methods, languages and tools that can be combined to let an entity specify, design, develop and test a CPS. 

Some problems were approached through the eyes of the Model checking methodology to verify the safety properties. Akella and McMillin \cite{Akella:2009} modeled a CPS as a Security Process Algebra. They used model checker, CoPs, to check the confidentially properties. Bu \emph{et al.}\cite{Bu:2011} analyzed CPS aspects using a statistical model checker. Highlight that in this approach the state space explosion also occurs like in classical model checking problems. 

Zhang \emph{et al.} \cite{Zhang:2017} present a model to verify the safety properties in mobile CPS based on a SAT-based model checking algorithm. The system was modeled as a Petri net and presented a lower memory consumption.

A combined model checking and a new version of PALS (physically asynchronous, logically synchronous) were applied to verified an airplane turning control system in \cite{Bae:2015} and \cite{Bae:2016}. 
Bae~\emph{et al.}~\cite{Bae:2016} also present other applications as a networked thermostat controllers and networked water tank controllers with gravity component .

This paper takes advantage of one possible set of such techniques to model \emph{(LHA)}, develop \emph{(Maude)} and model check \emph{(Maude)} the leaking reservoir problem specified in \cite{reservoiarprobelm}. Additional formalisms that are capable of modeling the reservoir problem are presented in \cite{VPFormLanguages}. Some examples are: \emph{Differential Equations}, \emph{additional State Machines model beyond LHA}, \emph{Dataflow}, \emph{Discrete Events}. Each of these methods can be used alongside a plethora of tools and languages to develop and model check phases of a CPS.

Each formalism technique is going to favor a set of tools and languages when development moves to the next phase. The LHA approach used in this paper favors a model check approach such as \emph{Maude}. Of course other checkers such as \emph{SAL} \cite{SAL}, \emph{Spin} \cite{spin}, \emph{NuSMV} \cite{NUMSMV} and \emph{UPPAAL} \cite{UPPAAl} are viable options for the design process of a LHA based CPS. For an approach based on Discrete Event techniques, the language and tools would most likely been based on Hard Description Languages based on \emph{VHDL} like \emph{Verilog} \cite{VHDLandVerilog} or \emph{AMS with added extensions} \cite{VHDLandASM}. Meanwhile, a DataFlow based approach is more likely to leverage a language like \emph{Lustre} \cite{Lustre}. Each bringing their own set advantages and limitations, making it even harder to create consensus on a standardized end-to-to process to specify, validate and develop a CPS.

This situation highlights the toughest aspect of CPS design and implementation. Depending on the context of the project sponsors, their perspective can influence the formalism path adopted and push the project toward one of another set of languages and tools. This is an area of CPS design that could be improved through increased availability of references that provide a study case of such methods applied to the same problem, and provide a qualitative and quantitative comparison of the results of each individual implementation. Such works seem to be at a preliminary stage at this point in time.

\section{Conclusion}\label{sec:conclusion}

This work presents a first step in creating a formal model to modularly specify and model check CPS by precisely describing the concept of modular CPS modeling and illustrating it with a case study on how to model a CPS as a LHA and effectively implement it in RTM. It creates succinct code to simulate and model check the problem in this study scope.

We are currently developing an extension of RTM to support the notion of synchronous product of real-time systems as described in this paper. Our results are encouraging, as illustrated in this paper.

There are also opens further questions, which require further studies in terms of creating more precise models for the leaking reservoir models and other types of automata that can be used as starting point for CPS modeling alongside their respective implementation in RTM.

\section*{Acknowledgment}

This work was developed with the support of CAPES - Coordena\c{c}\~ao de Aperfei\c{c}oamento de Pessoal de N\'ivel Superior (Coordination for Enhancement of Higher Education Personnel, in Brazil) and FAPERJ - Funda\c{c}\~ao de Amparo a Pesquisa do Rio de Janeiro.


\end{document}